\documentclass[%
 reprint,
 superscriptaddress,
 amsmath,amssymb,
 aps,
 pra,
]{revtex4-2}

\usepackage[dvipsnames]{xcolor}
\usepackage{physics}
\usepackage{pgfplots}
\usepackage{float}
\usepackage{graphicx}
\usepackage{dcolumn}
\usepackage{bm}
\usepackage{tikz}
\usepackage{booktabs}
\usetikzlibrary{matrix, decorations.pathreplacing, positioning,quantikz2}

\definecolor{C1}{RGB}{52, 89, 149}
\definecolor{C2}{RGB}{251, 77, 61}
\definecolor{C3}{RGB}{3, 206, 164}
\definecolor{C4}{RGB}{202, 21, 81}
\usepackage{hyperref}
\hypersetup{colorlinks=true, linkcolor=C2, citecolor=C2, urlcolor=C2}

\usepackage[normalem]{ulem}

\newlength\figureheight
\newlength\figurewidth

\begin{document}

\title{Non-Clifford Crosstalk Noise in Surface Codes Using Hybrid Stabilizer-Tensor Network Methods}

\author{Ben Harper}
\affiliation{School of Physics, The University of Melbourne, Parkville, Victoria 3010, Australia}
\affiliation{Data61, CSIRO, Clayton, Victoria 3168, Australia}
\author{Azar C. Nakhl}
\affiliation{School of Physics, The University of Melbourne, Parkville, Victoria 3010, Australia}
\author{Martin Sevior}
\affiliation{School of Physics, The University of Melbourne, Parkville, Victoria 3010, Australia}
\author{Muhammad Usman}
\affiliation{School of Physics, The University of Melbourne, Parkville, Victoria 3010, Australia}
\affiliation{Data61, CSIRO, Clayton, Victoria 3168, Australia}
\affiliation{School of Physics \& Astronomy, Monash University, Clayton, VIC 3800, Australia}

\begin{abstract}
Scalable realisation of quantum computing is reliant on the development of fault tolerant devices. Analysis of quantum error correction protocols typically considers incoherent noise models or noise-free syndrome measurements. While this is simple to simulate classically and straightforward to compute analytically, these simplifications are unable to capture the full dynamics of a noisy quantum system. In this work we use advanced hybrid stabilizer-tensor network simulation techniques to simulate coherent quantum crosstalk noise during syndrome extraction on a surface code. We show that the inclusion of coherence increases logical error rates and lowers the code threshold. In addition, we show that the specific distribution of the noise can quantitatively change logical error rates. The methods in this work allow simulation of quantum error correction with noise models previously inaccessible to classical simulation, providing new insights on the effect of crosstalk noise on quantum error correction codes.
\end{abstract}

\maketitle

\section{Introduction}
Quantum computing has advanced rapidly in recent years, with experimental platforms demonstrating increasing qubit counts, improved coherence times, and steadily decreasing physical error rates~\cite{google_quantum_ai_and_collaborators_quantum_2025,quantinuum_helios_2025,mandelbaum_scaling_nodate}. As a result, several leading architectures are approaching the regime in which fault-tolerant quantum computation—enabled by quantum error correction (QEC) may become feasible within the next few years. In QEC, logical information is encoded across many physical qubits and protected through repeated rounds of syndrome extraction and correction~\cite{shor_scheme_1995}. A central feature of this framework is the existence of a threshold: when physical error rates fall below this value, increasing the code distance suppresses logical errors exponentially~\cite{knill_resilient_1998}. Recent experimental work~\cite{google_quantum_ai_and_collaborators_quantum_2025} has shown error suppression as a function of code distance, which is an early indication of experimental quantum error correction. In this near-threshold regime, a quantitative understanding of how realistic noise processes affect large-scale error correction circuits is increasingly important, as small deviations from idealized assumptions can significantly alter logical error behavior, even breaking down the threshold theorem of surface codes~\cite{f_kam_detrimental_2025}.

Simulation of QEC circuits involves large numbers of entangled qubits. As such, a variety of approximations are employed such as simulating only up to distance 5 codes~\cite{huang_alibaba_2020,hakkaku_sampling-based_2021,obrien_density-matrix_2017,manabe_efficient_2025}, noise free syndrome extraction circuits~\cite{bravyi_correcting_2018,darmawan_tensor-network_2017}, or approximating coherent noise by the Pauli twirling approximation~\cite{zhou_surface_2025} (PTA), where coherent errors are replaced a stochastic Pauli error derived by Pauli twirling the coherent noise~\cite{katabarwa_logical_2015}. While these approaches provide valuable insights, they are subject to finite size effects due to being dominated by the boundary, or ignore features such as coherent, spatial correlations and multi-qubit crosstalk. Incorporating such effects into circuit-level simulations has historically been limited by the exponential cost of simulating non-Clifford dynamies at scale. There has been recent interest in overcoming these limitations for local coherent noise~\cite{behrends_surface_2025,leblond_logical_2025}.

An important source of noise in quantum computers is crosstalk noise, which is a broad term for any kind of unintended interactions between qubits, leading to correlated errors~\cite{gicev_crosstalk_2026}. This may be due to a variety of physical processes depending on the hardware platform in question. Recent work has shown that crosstalk noise may be particularly detrimental to surface codes~\cite{catelani_quantum_2026,zhou_surface_2025}, however these existing works have focused on a stochastic approximation of coherent effects, which does not capture the full dynamics of coherent noise.

Large quantum error correction circuits are typically simulated using stabilizer simulators such as Stim~\cite{gidney_stim_2021}, as they can efficiently simulate circuits that are composed exclusively from Clifford gates, as QEC circuits are. In contrast, tensor network simulation methods~\cite{orus_practical_2014} are able to simulate circuits with low entanglement, making them a poor choice for QEC circuits, despite their advantage of supporting arbitrary quantum operations. Recent advances in classical simulation techniques have unified these methods to create hybrid stabilizer tensor network simulation methods~\cite{harper_gcamps_2026,nakhl_stabilizer_2025,masot-llima_stabilizer_2024} which are able to leverage the advantages of both simulation methods. This presents an opportunity to use the new hybrid methods to simulate quantum error correction under noise models that were previously inaccessible. In parallel to this work, simulation methods based on ZX-calculus~\cite{kissinger_simulating_2022,haenel_tsim_2026} also allow simulating mostly Clifford circuits efficiently.

In this work, we apply the advanced hybrid simulation library GCAMPS~\cite{harper_gcamps_2026,nakhl_stabilizer_2025} to perform large-scale circuit-level simulations of the rotated surface code subject to coherent non-Clifford crosstalk noise. This enables the direct numerical study of noise models that were previously impractical to simulate, without resorting to the approximations used by previous studies of crosstalk on the surface code~\cite{huang_alibaba_2020,zhou_surface_2025}. We show that while coherent simulation does not produce significantly different thresholds to a PTA of crosstalk noise, there is a substantial increase in logical error rates for physical error rates below threshold. We also simulate different crosstalk noise models with the same Pauli twirled approximation, and find differences in logical error rate that would not be captured in a standard PTA simulation. These results highlight the critical importance of considering the coherent effects of crosstalk in surface code simulation, while also presenting valuable methods for further research of non-Clifford effects on the surface code.

\section{Surface Code Error Correction}
\begin{figure*}
    \centering
    \begin{minipage}{\linewidth}
        \begin{minipage}{0.03\linewidth}
            (a)
        \end{minipage}
        \begin{minipage}{0.23\linewidth}
            \includegraphics[width=1.0\linewidth]{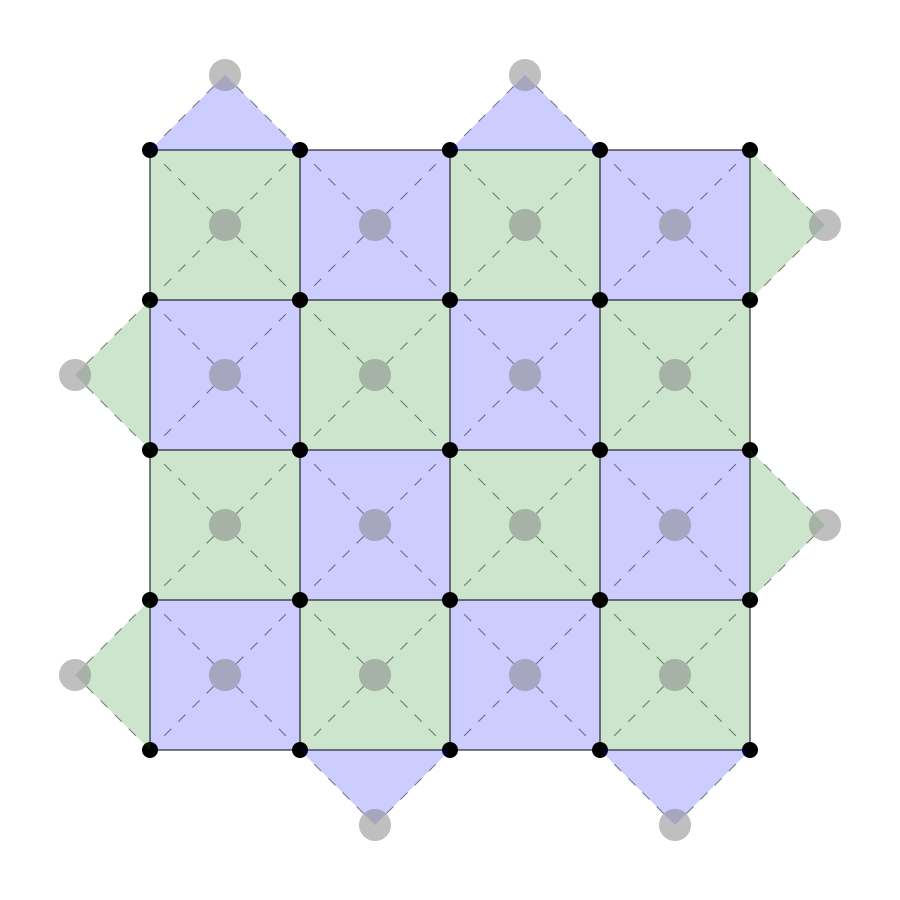}
        \end{minipage}
        \begin{minipage}{0.03\linewidth}
            (b)
        \end{minipage}
        \begin{minipage}{0.3\linewidth}
            \begin{quantikz}[row sep=-1pt, column sep=0.2cm]
                \lstick{$\ket{0}$} & \hphantomgate{H} & \targ{} \slice{} & \targ{} \slice{} & \targ{} \slice{} & \targ{} \slice{} & \hphantomgate{H} & \meter{} \\
                \lstick{$\ket{d_1}$} & & \ctrl{-1} & & & & \ghost{X} & \\
                \lstick{$\ket{d_2}$} & & & \ctrl{-2} & & & \ghost{X} & \\
                \lstick{$\ket{d_3}$} & & & & \ctrl{-3} & & \ghost{X} & \\
                \lstick{$\ket{d_4}$} & & & & & \ctrl{-4} & \ghost{X} &
            \end{quantikz}
            \centering
            $Z$ type stabilizer
        \end{minipage}
        \begin{minipage}{0.03\linewidth}
            (c)
        \end{minipage}
        \begin{minipage}{0.3\linewidth}
            \begin{quantikz}[row sep=-1pt, column sep=0.2cm]
                \lstick{$\ket{0}$} & \gate{H} & \ctrl{1} \slice{} & \ctrl{2} \slice{} & \ctrl{3} \slice{} & \ctrl{4} \slice{} & \gate{H} & \meter{} \\
                \lstick{$\ket{d_1}$} & & \targ{} & & & & & \ghost{X} \\
                \lstick{$\ket{d_2}$} & & & \targ{} & & & & \ghost{X} \\
                \lstick{$\ket{d_3}$} & & & & \targ{} & & & \ghost{X} \\
                \lstick{$\ket{d_4}$} & & & & & \targ{} & & \ghost{X}
            \end{quantikz}
            \centering
            $X$ type stabilizer
        \end{minipage}
    \end{minipage}
    \caption{(a) A distance 5 rotated surface code. Data qubits ($\bullet$) are on vertices of the grid, while ancilla qubits (\textcolor{lightgray}{$\bullet$}) are on the faces. Each square face measures alternating $X$ and $Z$ type stabilizers. Stabilizers on the boundary include only two data qubits.
    Circuits for (b) $Z$ and (c) $X$ syndrome extraction on the surface code. The ancilla is prepared in the $\ket{0}$ state and measures $Z_1 Z_2 Z_3 Z_4$ or $X_1 X_2 X_3 X_4$ on the four adjacent data qubits $d_i$. Vertical dashed lines indicate when crosstalk noise occurs after the entangling CNOT gates in each syndrome extraction circuit.}\label{fig:surface-code}
\end{figure*}

Quantum error correction is essential for the practical implementation of large scale quantum circuits. The rotated surface code~\cite{bombin_optimal_2007} is a topological stabilizer code with local parity check operators that allow efficient detection of errors through repeated measurement. Data qubits are arranged on the vertices of a square lattice, as shown in Figure~\ref{fig:surface-code}(a), while ancilla qubits are located on the faces of the lattice. During error correction, the stabilizers $X_1 X_2 X_3 X_4$ and $Z_1 Z_2 Z_3 Z_4$ associated with each face are extracted using the ancilla qubits and the circuits shown in Figure~\ref{fig:surface-code}(b) and (c). This process is repeated for $d$ rounds, producing ancilla measurement outcomes known as the error syndrome. These are processed by a classical decoder~\cite{higgott_sparse_2025,gicev_fully_2026,ott_decision-tree_2025} which infers a likely physical error consistent with the syndrome, and returns the corresponding correction. If this process fails the resulting logical state will be incorrect. The rate of logical error as a function of the physical error rate is a key result in quantum error correction work.

When considering coherent physical noise, the resulting logical state may experience a coherent non-Pauli error~\cite{bravyi_correcting_2018}. As such, it is useful to define the logical error rate $P_L$,
\begin{equation}
    P_L = \frac{1}{N} \sum_{i=1}^N |\sin(\theta_i/2)|,\label{eq:P_L}
\end{equation}
where we sum over the logical rotation angles $\theta_i$ resulting from $N$ error syndromes sampled from the probability distribution of our noise model. This is the average diamond norm distance between the logical error channel and the identity channel~\cite{bravyi_correcting_2018} and serves as a convenient proxy for logical error in the presence of coherent noise. In the case of Pauli errors $\theta_i \in \{0, \pi\}$ and so Equation~\ref{eq:P_L} reduces to the standard logical error rate.

\section{Noise Models}\label{sec:noise-models}
Noise in quantum devices can take many different forms, such as depolarising, dephasing, relaxation, leakage and more. When analysing quantum error correction codes numerically, this noise is typically represented by stochastic Pauli operations. This is because QEC circuits contain only Clifford gates, and if a noise model is also Clifford, then circuits may be simulated efficiently classically using a stabilizer simulator~\cite{aaronson_improved_2004} such as Stim~\cite{gidney_stim_2021}. This setup however is unable to capture the full dynamics of a coherent noise model. The difficulty in simulation has lead to previous studies of noise beyond Pauli models on the surface code have been limited to simplifications such as perfect syndrome extraction circuits~\cite{bravyi_correcting_2018} or a Pauli-twirled approximation of the noise~\cite{zhou_surface_2025}.

\subsection{Baseline Depolarizing}
Simulations in this paper consist of a noise model with a baseline depolarizing component and a crosstalk component. The depolarizing component consists of an incoherent Pauli gate chosen at random and inserted after each physical gate with probability $p_1$ for single qubit gates and $p_2$ for two qubit gates,
\begin{align}
    \varepsilon_1(\rho) &= (1 - p_1) \rho + \frac{p_1}{3} \sum_i \sigma_i \rho \sigma_i, \\
    \varepsilon_2(\rho) &= (1 - p_2) \rho + \frac{p_2}{15} \sum_{i,j} (\sigma_i \otimes \sigma_j) \rho (\sigma_i \otimes \sigma_j),
\end{align}
where the sums are over the set of Pauli operators where $i, j \in \{I, X, Y, Z\}$ and $(i, j) \neq (I, I)$

We parameterise this noise by a single parameter $p$ which is scaled according to gate type based on the observations that different physical gates experience different error rates. In particular, two-qubit gates are significantly more noisy than single qubit gates, and measurements are noisier again. Table~\ref{tab:error-rates} shows the error rate ratios used in this work. The parameter $p$ is then varied to find the code's threshold, the point where increasing the code distance suppresses logical error rates.

\begin{table}
\centering
\begin{tabular}{cc}
\toprule
Noise Model & Error Rate \\
\midrule
Single qubit Clifford Gate & $p_1 = 0.1 p$ \\
Two qubit Clifford Gate & $p_2 = p$ \\
Reset & $p_R = 2p$ \\
Measurement & $p_M = 5p$ \\
Crosstalk & $\theta = 10^{-3}$ \\
\bottomrule
\end{tabular}
\caption{Error rates used in this paper. The parameter $p$ is varied to find the threshold, while the strength of the crosstalk parameter $\theta$ is fixed.}\label{tab:error-rates}
\end{table}

\subsection{Gate-based Crosstalk}
The model of crosstalk considered in this paper is gate based nearest neighbour, that is, crosstalk noise which occurs between nearest neighbour qubits when a gate is performed on the quantum computer. This is commonly observed in tuneable transmon superconducting qubits (such as in those produced by Google~\cite{google_quantum_ai_and_collaborators_quantum_2025}), where the two qubit interaction is supressed by detuning when a gate is not being applied. A related source of crosstalk is an always on interaction, present in devices with fixed transmons (e.g. IBM quantum hardware~\cite{sundaresan_reducing_2020, wesdorp_mitigating_2026}). We note however that in the case of QEC circuits these two distinct modes are qualitatively similar, as QEC circuits consist almost entirely of two qubit gates and so qubits are almost always interacting. In this context the distinction between these two noise channels is determined by the precise noise parameters of the hardware, which is a detail beyond the scope of this work.

We model crosstalk as a coherent $ZZ$ rotation between nearest neighbours when a two qubit gate is applied, as shown in Figure~\ref{fig:surface-code}. The noise channel is,
\begin{equation}
    \varepsilon(\rho) = e^{i\theta \sigma_{Z1} \otimes \sigma_{Z2}} \rho e^{-i\theta \sigma_{Z1} \otimes \sigma_{Z2}},\label{eq:coherent-crosstalk}
\end{equation}
where the noise parameter $\theta$ is defined by the residual coupling strength $J_{ZZ}$ and the gate length $t_g$,
\begin{equation}
    \theta = J_{ZZ} t_g.
\end{equation}
These terms are hardware dependent, though typically $J_{ZZ}$ is around 100kHz while $t_g$ is of the order of 100ns~\cite{zhao_high-contrast_2020,kandala_demonstration_2021}. This results in a value of $\theta$ of the order of $10^{-3}$, which we use for the simulations in this work. The noise channel is implemented in the simulator using the following sequence of gates CNOT and single qubit $Z$ rotation gates:
\begin{equation*}
\begin{quantikz}
    & \ctrl{1} & & \ctrl{1} & \\
    & \targ{} & \gate{R_Z(\theta/2)} & \targ{} & 
\end{quantikz}
\end{equation*}

\subsection{Pauli Twirling Approximation}
A common approach used to approximate the effect of coherent noise is Pauli twirling; that is, replace the coherent noise with a stochastic Pauli noise model where the probability of a Pauli occurring is a function of the coherent rotation angle~\cite{zhou_surface_2025}. In the case of the coherent noise described in Equation~\ref{eq:coherent-crosstalk}, the Pauli twirling approximation (PTA) is,

\begin{align}
    \varepsilon_\text{twirl}(\rho) &= (1-\sin^2 (\theta)) \rho \nonumber \\
    &\quad + \sin^2(\theta) (\sigma_Z \otimes \sigma_Z) \rho (\sigma_Z \otimes \sigma_Z)\label{eq:pauli-twirling}
\end{align}

Mathematically, the PTA is a projection of the Pauli transfer matrix of a channel onto the diagonal. If the off diagonal terms are small then this is approximation is justified. For a coherent noise channel, this is the case when a measurement immediately collapses the error to the Pauli basis, however twirling removes phase information and during further evolution coherent errors can interfere between error correction rounds in a way that incoherent errors do not. To quantify the effect of coherence, we simulate both the fully coherent noise model and the Pauli twirled approximation for comparison.

\subsection{Error Detection}
Physical errors are identified from syndrome measurements using minimum weight perfect matching (MWPM) via the PyMatching~\cite{higgott_sparse_2025} Python package. As PyMatching only supports Pauli type error models, the error models used for all error correction in this work are generated from the Pauli Twirling Approximation.

\section{Simulation Method}
Simulation of general coherent noise on a surface code is extremely challenging. The large number of qubits present limits direct state vector simulation, while the global entanglement between data qubits prohibits simulation by tensor network methods. In this paper we use the recently developed hybrid stabilizer-tensor network simulation library GCAMPS~\cite{harper_gcamps_2026} to simulate the coherent noise models considered. Stabilizer tensor networks were recently introduced and are able to efficiently simulate circuits with large amounts of entanglement or non-Clifford gates, but not both.

\subsection{Stabilizer Tensor Networks}
In a stabilizer tensor network simulator, an arbitrary state $\ket{\psi}$ is represented by a Clifford operator $C$ and a tensor network (typically a matrix product state) $\ket{\text{MPS}}$:
\begin{equation}
    \ket{\psi} = C \ket{\text{MPS}}\label{camps}
\end{equation}
A Clifford gate $G$ updates the Clifford operator $C$ directly, 

\begin{align*}
    G \ket{\psi} &= G C \ket{MPS} \\
                 &= C' \ket{MPS}.
\end{align*}
A non-Clifford operation $T$ must first be decomposed into a sum of Pauli operations $P$ and commuted through $C$ before updating the MPS.
\begin{align*}
    U \ket{\psi} &= U C \ket{MPS} \\
                 &= \sum_i P_i C \ket{MPS} \\\
                 &= C \sum_i \tilde{P}_i \ket{MPS} \\\
                 &= C \ket{MPS'}
\end{align*}
Note in particular that what was a local Pauli string may now have higher weight (the number of non-identity terms in the string), depending on the entanglement present in $C$. The operator $C$ transforms physically local operations into non-local operations on the tensor network. Projective measurement is implemented in a similar way, where a sum of Paulis is commuted through $C$ and applied to the tensor network directly.

In the specific context of quantum error correction circuits, we can further interpret the components of $\ket{\psi}$. The operator $C$ is the ideal Clifford operator that implements the error correction code. When non-Clifford errors occur, they do not change the Clifford $C$. Hence the state of the MPS corresponds to the error that perturbs the ideal Clifford state captured in $C$. When a measurement is made, the non-Clifford error described in the MPS collapses to a Pauli error in the Clifford tableau.

We finally note that in this work we do not perform typical optimisations that are possible for stabilizer tensor networks, specifically magic state injection~\cite{nakhl_stabilizer_2025} or Clifford optimisation~\cite{qian_augmenting_2024,lami_quantum_2025}. In the case of magic state injection, the large number of non-Clifford gates in these circuits would require a large number of ancilla qubits prepared in a magic state, while the cost of Clifford optimisation outweighed the benefits of reducing the tensor network's bond dimension.

\subsection{Truncation}\label{sec:truncation}
\begin{figure}
    \centering
    \includegraphics[width=0.9\linewidth]{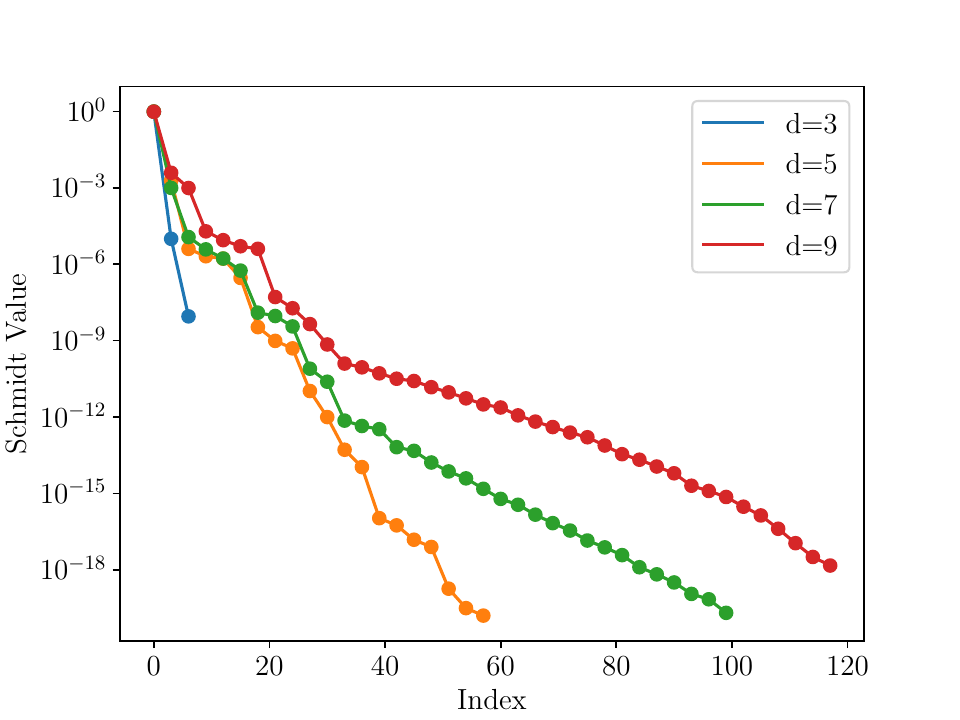}
    \caption{Schmidt values squared for a surface code simulation with $d=3, 5, 7, 9$. The rapid exponential decay in the Schmidt values shown here allows for substantial truncation of the tensor network without impairing its representation of the surface code state. For a code with $n$ physical qubits, the Schmidt values are across the cut between $n/2$ and $n/2 - 1$.}\label{fig:schmidt-values}
\end{figure}
\begin{figure}
    \includegraphics[width=0.9\linewidth]{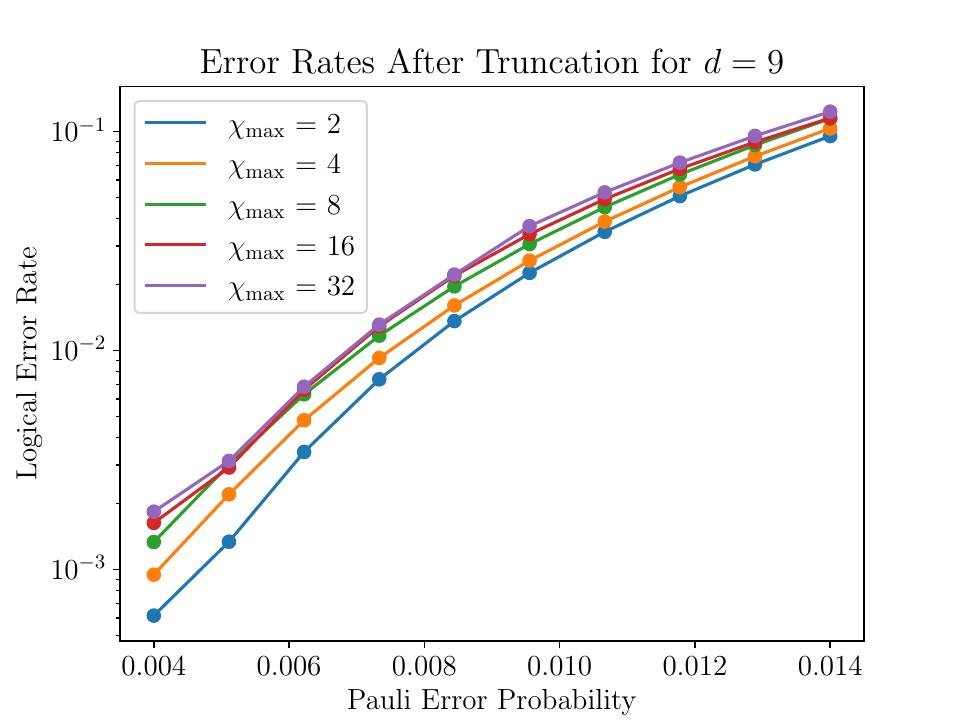}
    \caption{The logical error rate plotted for various levels of truncation for a distance $d=9$ surface code. Despite doubling the number of terms not truncated at each step, the measured error rates rapidly converge. Note that reducing the number of remaining terms reduces the logical error rate.}
    \label{fig:truncated-error-rate}
\end{figure}
To gather useful statistics from quantum error correction circuits, hundreds of thousands of samples are required. Even with a stabilizer tensor network simulator, this is a challenging task. To further improve the simulation performance, we limit the bond dimension of the tensor network to a maximum value, reduing the simulation cost at the expense of simulation accuracy. When performing a singular value decomposition on the tensor network, the state across the cut looks like,

\begin{equation}
    \ket{\psi} = \sum_{i=1}^{\chi-1} \lambda_i \ket{i_L} \ket{i_R},
\end{equation}
where $\ket{i_L}$ and $\ket{i_R}$ are the state to the left and right of the cut, while $\lambda_i$ are the singular values at the cut. Truncation involves limiting the bond dimension $\chi$ to $\chi_\text{max} \leq 2^{N/2}$ and throwing away terms in the decomposition. When the singular values are ordered by size, this means discarding the terms with the smallest contribution to the final state. This introduces imprecision in the simulation results, as it is no longer possible to reach the the resulting state in a truncated tensor network. As such, it is important to ensure that the truncation does not significantly change the expectation values and measurement statistics of interest. In the case of the error correction circuits we simulate here, that means that after truncation the observed logical error rates should be close to their true value.

To judge the effect of truncation, we perform simulation of the surface code with the crosstalk noise model described above. In Figure~\ref{fig:schmidt-values} we plot the square of the Schmidt values ($\lambda_i^2$) for a cut at the centre of the MPS. Note that the Schmidt values are proportional to the measurement probabilities of the state $\ket{\psi}$. These results show that for all distances considered, the Schmidt values decrease exponentially while the sum of the remaining Schmidt values is close to one, allowing substantial truncation of the tensor network.

Finally, in Figure~\ref{fig:truncated-error-rate} we show the logical error rate for $d=9$ for various levels of truncation. These results confirm the rapid convergence of the measured results and the futility of simulating larger tensor networks, validating truncation as an optimisation strategy. We note that when the truncation is too large, the measured logical error rates are lower. Intuitively, this is because the largest component of the MPS is the zero state corresponding to no crosstalk noise. The terms in the MPS with lower probability correspond to noise, and while the dynamics of QEC circuits at large distances can be determined by highly improbable events, the exponential decay in Schmidt values suppresses the effect on the final results. However, this does imply that our results are a lower bound on the logical error rates.

As such, all further Figures in this paper consist of simulations with the MPS truncated to a maximum bond of $\chi_\text{max} = 32$.

\section{Results}
\subsection{Crosstalk Noise During Syndrome Extraction}
\begin{figure*}
    \centering
    (a) \includegraphics[width=0.45\linewidth]{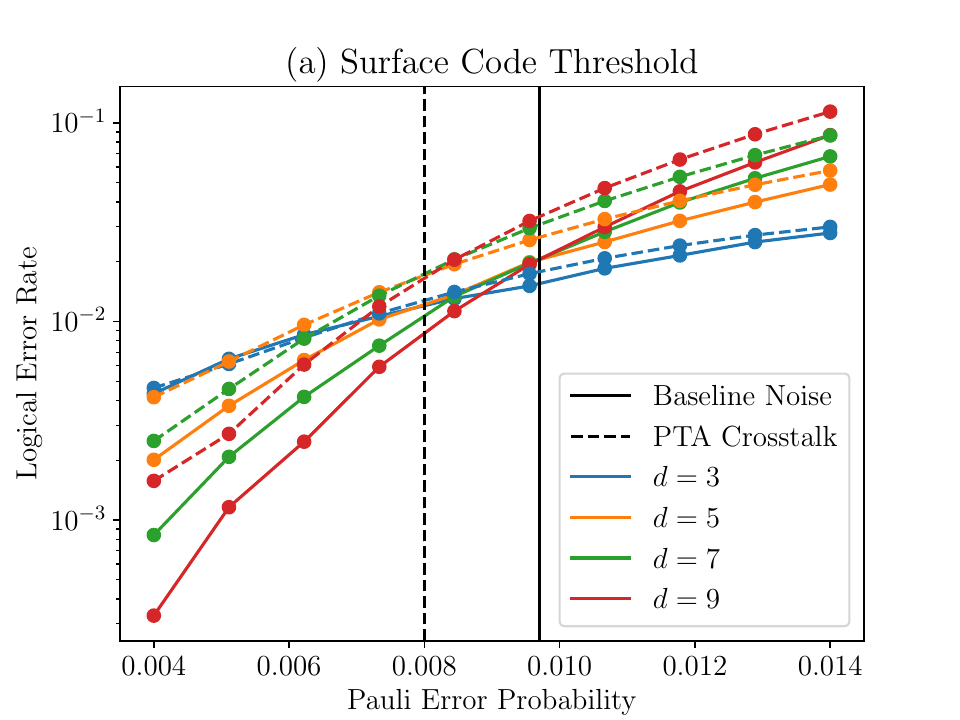}
    \includegraphics[width=0.45\linewidth]{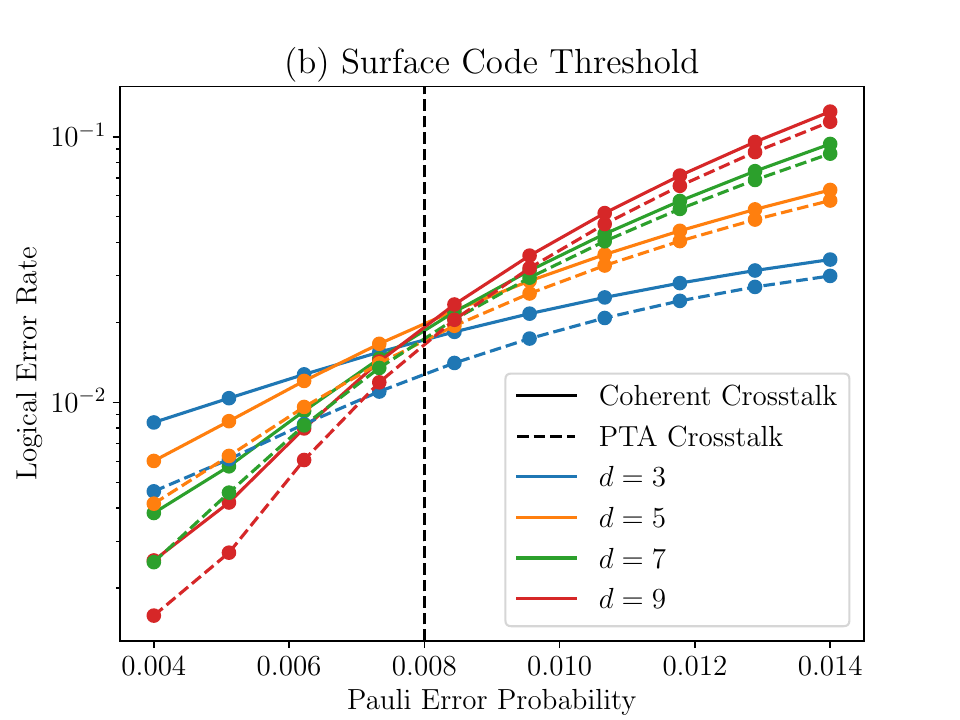}
    \caption{Logical error rates for the surface code under the baseline Pauli noise model with no crosstalk compared with (a) incoherent Pauli twirled crosstalk and (b) coherent crosstalk noise. The Pauli only noise model is the least detrimental to logical error rates, while the coherent crosstalk noise model is the most detrimental. The vertical lines indicate the threshold for each type of noise.}
    \label{fig:crosstalk-thresholds}
\end{figure*}

We now present the simulation results for the crosstalk noise during syndrome extraction described above in Section~\ref{sec:noise-models}. In these simulations we use a physically plausible $J_{ZZ} = 150\text{kHz}$ and $t_g = 150\text{ns}$. Figure~\ref{fig:crosstalk-thresholds} shows the measured logical error rates for Pauli twirled crosstalk compared to (a) the baseline noise model and (b)  coherent crosstalk noise. Each data point plotted is an average derived from $10^5$ samples. We first note that the inclusion of crosstalk increases logical error rates, and reduce the threshold from 1\% to 0.8\%. The further inclusion of coherence increases logical error rates again, though does not have a statistically significant effect on the code's threshold. While errors are still exponentially supressed below the threshold, a higher logical error rate for the code necessitates lower physical error rates or larger code distances to reach the same performance.

\subsection{Alternative Noise Distributions}
\begin{figure}
    \includegraphics[width=0.9\linewidth]{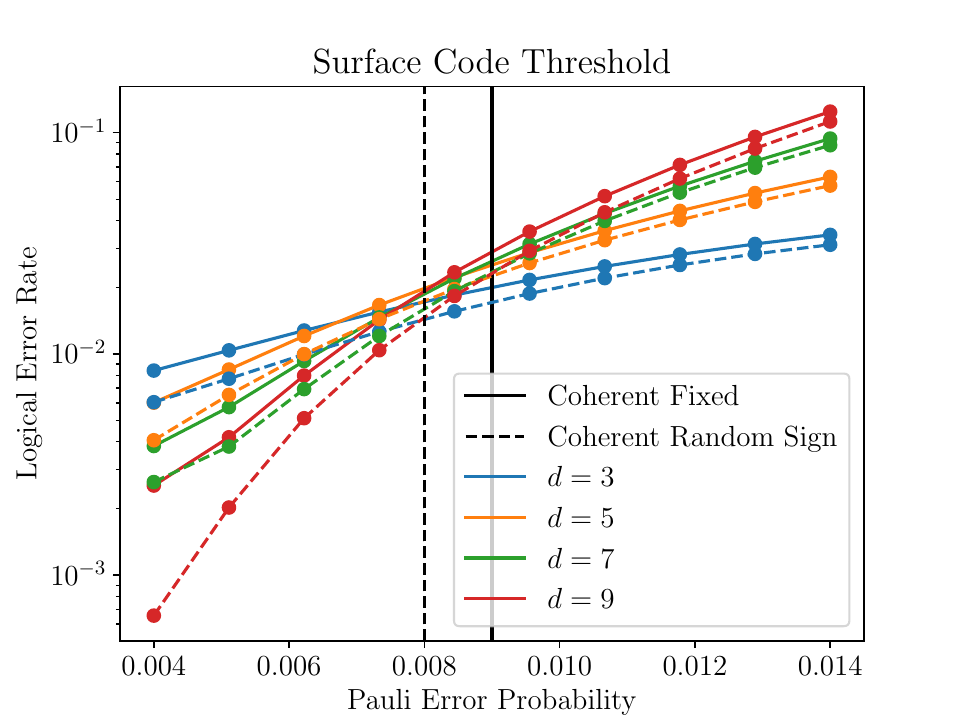}
    \caption{Logical error rates for the surface code where the sign of the coherent crosstalk rotation angle is chosen uniformly at random. The logical error rates are identical to the Pauli Twirling approximation, despite the coherence of the noise.}
    \label{fig:random-dir}
\end{figure}
\begin{figure}
    \centering
    \includegraphics[width=0.9\linewidth]{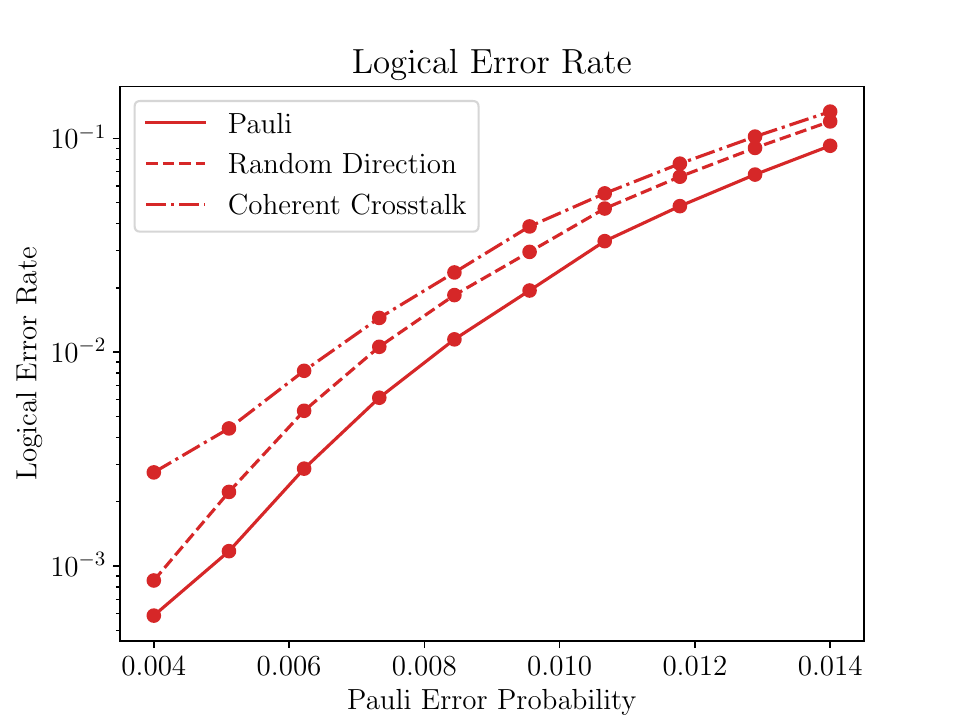}
    \caption{Logical error rates under each noise model for $d=9$. The inclusion of coherent crosstalk noise increases logical error rates over the baseline Pauli noise model, while coherent crosstalk noise in random directions somewhat reduces logical error rates again.}
    \label{fig:d9}
\end{figure}

To illustrate the effect of coherence on surface code thresholds, we now simulate an alternative shape for crosstalk noise from the uniform noise presented in the previous subsection. Note however that the noise considered here and in the previous section have identical Pauli twirled approximations, highlighting the importance of proper consideration of the coherent interference in noise. We consider the noise model where the angle $\theta$ is fixed as in the previous section, but its sign is chosen uniformly at random,

\begin{equation}
    \theta_i \in \{ \theta, -\theta \}.
\end{equation}

The Pauli twirling approximation of the physical noise is the same as in Equation~\ref{eq:pauli-twirling}, as $\sin^2(\theta) = \sin^2(-\theta)$. However while the inclusion of coherence in the original noise model resulted in constructive interference, in this case the noise interferes destructively. Figure~\ref{fig:random-dir} shows this noise model in addition to the fixed crosstalk noise from above. Interestingly, while the difference between the noise models is minimal above threshold, below the threshold the logical error rate of the random sign noise model is significantly below the fixed crosstalk noise model. This shows that simplifying coherent noise by its Pauli twirled approximation cannot capture the full dynamics of a noisy quantum system. Finally, in Figure~\ref{fig:d9} we show the logical error rates for each noise model on a distance 9 code. Despite the two coherent noise models under consideration having identical Pauli twirling approximations, Figure~\ref{fig:d9} demonstrates that they are inherently different. While physical error rates below the error correction threshold will result in exponential suppression of the logical error rate in either case, reaching a given level of logical error rate may be more difficult under some coherent noise models. It may be necessary to reduce physical error rates further below the threshold to obtain a target logical error rate, or increase code distances, reducing efficiency of the code. These effects can only captured in a coherent simulation of the surface code under noise.

\section{Conclusion}
In this work we present the application of stabilizer tensor network simulation to the problem of simulating coherent crosstalk noise during syndrome extraction on large error correction circuits. We find that quantum error correction still works well at reducing logical error rates once noise is below the threshold. However, when considering coherent errors during error correction circuits, logical error rates are higher than suggested by the Pauli twirling approximation, while the threshold is approximately 0.8\%. We also show that, perhaps unsurprisingly, different distributions of coherent noise with the same Pauli twirled approximation can show quantitatively different logical error rates and thresholds. In particular, crosstalk noise with a uniform coherent rotation angle is significantly more disruptive to logical error rates than coherent crosstalk noise in a randomly chosen direction. Proper consideration of coherent noise will be essential to the task of accurately characterising quantum hardware in preparation for fault tolerant quantum computing.

The methods presented here are a novel framework for simulating arbitrary coherent noise in surface codes in a practical amount of time. Further work may consider other noise models of interest to particular hardware platforms, such as amplitude damping, leakage and non-Markovian noise~\cite{white_demonstration_2020,figueroa-romero_randomized_2021}. Further optimisation of the simulation to reduce cost may be possible, such as some analytic simplification of noise models to reduce the number of non-Clifford gates that must be simulated. It may also be beneficial to use other tensor network layouts, such as Projected Entangled Pair States (PEPS)~\cite{lee_scalable_2025} or Tree Tensor Networks (TTNs)~\cite{seitz_simulating_2023}. These methods may be applied to the study of non-Pauli noise models on other quantum error correction codes, such as quantum low-density parity check (qLDPC) codes~\cite{bravyi_high-threshold_2024}. Finally, these methods may allow efficient calculation of computation thresholds, a task with is substantially more computationally involved than the memory experiments conducted here.

\section*{Acknowledgment}
BH acknowledges the support of
the CSIRO Research Training Program
Scholarship. ACN acknowledges the support of the Australian Government Research Training Program Scholarship. The work was partially
supported by the Australian Army through Quantum
Technology Challenge 2023. This research was supported by The University of Melbourne’s Research Computing Services and the Petascale Campus Initiative. The research was supported by the University of Melbourne through the establishment of the IBM Quantum Network Hub at the University. \\
\textbf{Data availability:} The data that support the findings of this study are available within the article. \\
\textbf{Competing financial interests:} The authors declare no competing financial or non-financial interests.

\bibliography{references}

@article{aaronson_improved_2004,
  title = {Improved Simulation of Stabilizer Circuits},
  author = {Aaronson, Scott and Gottesman, Daniel},
  year = 2004,
  journal = {Physical Review A},
  volume = {70},
  number = {5},
  pages = {052328},
  publisher = {APS},
  doi = {10.1103/PhysRevA.70.052328}
}

@article{behrends_surface_2025,
  title = {The Surface Code beyond {{Pauli}} Channels: {{Logical}} Noise Coherence, Information-Theoretic Measures, and Errorfield-Double Phenomenology},
  shorttitle = {The Surface Code beyond {{Pauli}} Channels},
  author = {Behrends, Jan and B{\'e}ri, Benjamin},
  year = 2025,
  month = dec,
  journal = {PRX Quantum},
  volume = {6},
  number = {4},
  eprint = {2412.21055},
  primaryclass = {quant-ph},
  pages = {040350},
  issn = {2691-3399},
  doi = {10.1103/psf5-b6j2},
  urldate = {2026-03-17},
  archiveprefix = {arXiv}
}

@article{bombin_optimal_2007,
  title = {Optimal Resources for Topological Two-Dimensional Stabilizer Codes: {{Comparative}} Study},
  shorttitle = {Optimal Resources for Topological Two-Dimensional Stabilizer Codes},
  author = {Bombin, H. and {Martin-Delgado}, M. A.},
  year = 2007,
  month = jul,
  journal = {Physical Review A},
  volume = {76},
  number = {1},
  pages = {012305},
  publisher = {American Physical Society},
  doi = {10.1103/PhysRevA.76.012305},
  urldate = {2026-05-05}
}

@article{bravyi_correcting_2018,
  title = {Correcting Coherent Errors with Surface Codes},
  author = {Bravyi, Sergey and Englbrecht, Matthias and K{\"o}nig, Robert and Peard, Nolan},
  year = 2018,
  month = oct,
  journal = {npj Quantum Information},
  volume = {4},
  number = {1},
  pages = {55},
  publisher = {Nature Publishing Group},
  issn = {2056-6387},
  doi = {10.1038/s41534-018-0106-y},
  urldate = {2025-12-17},
  copyright = {2018 The Author(s)},
  langid = {english}
}

@article{bravyi_high-threshold_2024,
  title = {High-Threshold and Low-Overhead Fault-Tolerant Quantum Memory},
  author = {Bravyi, Sergey and Cross, Andrew W. and Gambetta, Jay M. and Maslov, Dmitri and Rall, Patrick and Yoder, Theodore J.},
  year = 2024,
  month = mar,
  journal = {Nature},
  volume = {627},
  number = {8005},
  pages = {778--782},
  publisher = {Nature Publishing Group},
  issn = {1476-4687},
  doi = {10.1038/s41586-024-07107-7},
  urldate = {2026-05-23},
  copyright = {2024 The Author(s)},
  langid = {english}
}

@article{catelani_quantum_2026,
  title = {Quantum {{Error Correction Faces Another Hurdle}}},
  author = {Catelani, Gianluigi},
  year = 2026,
  month = may,
  journal = {Physics},
  volume = {19},
  pages = {62},
  issn = {1943-2879},
  doi = {10.1103/Physics.19.62},
  urldate = {2026-05-05},
  langid = {english}
}

@article{darmawan_tensor-network_2017,
  title = {Tensor-{{Network Simulations}} of the {{Surface Code}} under {{Realistic Noise}}},
  author = {Darmawan, Andrew S. and Poulin, David},
  year = 2017,
  month = jul,
  journal = {Physical Review Letters},
  volume = {119},
  number = {4},
  pages = {040502},
  publisher = {American Physical Society},
  doi = {10.1103/PhysRevLett.119.040502},
  urldate = {2026-03-23}
}

@article{f_kam_detrimental_2025,
  title = {Detrimental Non-{{Markovian}} Errors for Surface Code Memory},
  author = {F Kam, John and Gicev, Spiro and Modi, Kavan and Southwell, Angus and Usman, Muhammad},
  year = 2025,
  month = jul,
  journal = {Quantum Science and Technology},
  volume = {10},
  number = {3},
  pages = {035060},
  publisher = {IOP Publishing},
  issn = {2058-9565},
  doi = {10.1088/2058-9565/adebab},
  urldate = {2026-04-21},
  langid = {english}
}

@article{figueroa-romero_randomized_2021,
  title = {Randomized {{Benchmarking}} for {{Non-Markovian Noise}}},
  author = {{Figueroa-Romero}, Pedro and Modi, Kavan and Harris, Robert J. and Stace, Thomas M. and Hsieh, Min-Hsiu},
  year = 2021,
  month = dec,
  journal = {PRX Quantum},
  volume = {2},
  number = {4},
  pages = {040351},
  publisher = {American Physical Society},
  doi = {10.1103/PRXQuantum.2.040351},
  urldate = {2026-05-27}
}

@misc{gicev_crosstalk_2026,
  title = {Crosstalk {{In Contemporary Quantum Devices}}},
  author = {Gicev, Spiro and Harper, Ben and Kang, Haiyue and Usman, Muhammad and Sevior, Martin},
  year = 2026,
  month = may,
  number = {arXiv:2605.26528},
  eprint = {2605.26528},
  primaryclass = {quant-ph},
  publisher = {arXiv},
  doi = {10.48550/arXiv.2605.26528},
  urldate = {2026-05-27},
  archiveprefix = {arXiv}
}

@article{gicev_fully_2026,
  title = {Fully Convolutional {{3D}} Neural Network Decoders for Surface Codes with Syndrome Circuit Noise},
  author = {Gicev, Spiro and Hollenberg, Lloyd and Usman, Muhammad},
  year = 2026,
  month = apr,
  journal = {Quantum Science and Technology},
  issn = {2058-9565},
  doi = {10.1088/2058-9565/ae5fc9},
  urldate = {2026-05-06},
  copyright = {https://creativecommons.org/licenses/by/4.0/}
}

@article{gidney_stim_2021,
  title = {Stim: A Fast Stabilizer Circuit Simulator},
  author = {Gidney, Craig},
  year = 2021,
  journal = {Quantum},
  volume = {5},
  pages = {497},
  publisher = {Verein zur F\"orderung des Open Access Publizierens in den Quantenwissenschaften},
  doi = {10.22331/q-2021-07-06-497}
}

@article{google_quantum_ai_and_collaborators_quantum_2025,
  title = {Quantum Error Correction below the Surface Code Threshold},
  author = {{Google Quantum AI and Collaborators}},
  year = 2025,
  month = feb,
  journal = {Nature},
  volume = {638},
  number = {8052},
  pages = {920--926},
  publisher = {Nature Publishing Group},
  issn = {1476-4687},
  doi = {10.1038/s41586-024-08449-y},
  urldate = {2026-04-12},
  copyright = {2024 The Author(s)},
  langid = {english}
}

@misc{haenel_tsim_2026,
  title = {Tsim: {{Fast Universal Simulator}} for {{Quantum Error Correction}}},
  shorttitle = {Tsim},
  author = {Haenel, Rafael and Luo, Xiuzhe and Zhao, Chen},
  year = 2026,
  month = apr,
  number = {arXiv:2604.01059},
  eprint = {2604.01059},
  primaryclass = {quant-ph},
  publisher = {arXiv},
  doi = {10.48550/arXiv.2604.01059},
  urldate = {2026-05-27},
  archiveprefix = {arXiv}
}

@article{hakkaku_sampling-based_2021,
  title = {Sampling-Based Quasiprobability Simulation for Fault-Tolerant Quantum Error Correction on the Surface Codes under Coherent Noise},
  author = {Hakkaku, Shigeo and Mitarai, Kosuke and Fujii, Keisuke},
  year = 2021,
  month = nov,
  journal = {Physical Review Research},
  volume = {3},
  number = {4},
  pages = {043130},
  publisher = {American Physical Society},
  doi = {10.1103/PhysRevResearch.3.043130},
  urldate = {2026-04-17}
}

@inproceedings{harper_gcamps_2026,
  title = {{{GCAMPS}}: {{A Scalable Classical Simulator}} for {{Qudit Systems}}},
  shorttitle = {{{GCAMPS}}},
  booktitle = {Proceedings of the {{Supercomputing Asia}} and {{International Conference}} on {{High Performance Computing}} in {{Asia Pacific Region}}},
  author = {Harper, Ben and Nakhl, Azar and Quella, Thomas and Sevior, Martin and Usman, Muhammad},
  year = 2026,
  month = jan,
  series = {{{SCA}}/{{HPCAsia}} '26},
  pages = {1--9},
  publisher = {Association for Computing Machinery},
  address = {New York, NY, USA},
  doi = {10.1145/3773656.3773689},
  urldate = {2026-04-09},
  isbn = {979-8-4007-2067-3}
}

@article{higgott_sparse_2025,
  title = {Sparse {{Blossom}}: Correcting a Million Errors per Core Second with Minimum-Weight Matching},
  shorttitle = {Sparse {{Blossom}}},
  author = {Higgott, Oscar and Gidney, Craig},
  year = 2025,
  month = jan,
  journal = {Quantum},
  volume = {9},
  pages = {1600},
  publisher = {Verein zur F\"orderung des Open Access Publizierens in den Quantenwissenschaften},
  doi = {10.22331/q-2025-01-20-1600},
  urldate = {2026-04-10},
  langid = {british}
}

@misc{huang_alibaba_2020,
  title = {Alibaba {{Cloud Quantum Development Platform}}: {{Surface Code Simulations}} with {{Crosstalk}}},
  shorttitle = {Alibaba {{Cloud Quantum Development Platform}}},
  author = {Huang, Cupjin and Ni, Xiaotong and Zhang, Fang and Newman, Michael and Ding, Dawei and Gao, Xun and Wang, Tenghui and Zhao, Hui-Hai and Wu, Feng and Zhang, Gengyan and Deng, Chunqing and Ku, Hsiang-Sheng and Chen, Jianxin and Shi, Yaoyun},
  year = 2020,
  month = feb,
  number = {arXiv:2002.08918},
  eprint = {2002.08918},
  primaryclass = {quant-ph},
  publisher = {arXiv},
  doi = {10.48550/arXiv.2002.08918},
  urldate = {2026-04-17},
  archiveprefix = {arXiv}
}

@article{kandala_demonstration_2021,
  title = {Demonstration of a {{High-Fidelity}} Cnot {{Gate}} for {{Fixed-Frequency Transmons}} with {{Engineered}} \${{ZZ}}\$ {{Suppression}}},
  author = {Kandala, A. and Wei, K. X. and Srinivasan, S. and Magesan, E. and Carnevale, S. and Keefe, G. A. and Klaus, D. and Dial, O. and McKay, D. C.},
  year = 2021,
  month = sep,
  journal = {Physical Review Letters},
  volume = {127},
  number = {13},
  pages = {130501},
  publisher = {American Physical Society},
  doi = {10.1103/PhysRevLett.127.130501},
  urldate = {2026-05-07}
}

@article{katabarwa_logical_2015,
  title = {Logical Error Rate in the {{Pauli}} Twirling Approximation},
  author = {Katabarwa, Amara and Geller, Michael R.},
  year = 2015,
  month = sep,
  journal = {Scientific Reports},
  volume = {5},
  number = {1},
  pages = {14670},
  publisher = {Nature Publishing Group},
  issn = {2045-2322},
  doi = {10.1038/srep14670},
  urldate = {2026-05-23},
  copyright = {2015 The Author(s)},
  langid = {english}
}

@article{kissinger_simulating_2022,
  title = {Simulating Quantum Circuits with {{ZX-calculus}} Reduced Stabiliser Decompositions},
  author = {Kissinger, Aleks and {van de Wetering}, John},
  year = 2022,
  month = jul,
  journal = {Quantum Science and Technology},
  volume = {7},
  number = {4},
  pages = {044001},
  publisher = {IOP Publishing},
  issn = {2058-9565},
  doi = {10.1088/2058-9565/ac5d20},
  urldate = {2026-05-27},
  langid = {english}
}

@article{knill_resilient_1998,
  title = {Resilient {{Quantum Computation}}},
  author = {Knill, Emanuel and Laflamme, Raymond and Zurek, Wojciech H.},
  year = 1998,
  month = jan,
  journal = {Science},
  volume = {279},
  number = {5349},
  pages = {342--345},
  issn = {0036-8075, 1095-9203},
  doi = {10.1126/science.279.5349.342},
  urldate = {2026-05-05},
  langid = {english}
}

@article{lami_quantum_2025,
  title = {Quantum {{State Designs}} with {{Clifford-Enhanced Matrix Product States}}},
  author = {Lami, Guglielmo and Haug, Tobias and De Nardis, Jacopo},
  year = 2025,
  month = mar,
  journal = {PRX Quantum},
  volume = {6},
  number = {1},
  pages = {010345},
  publisher = {American Physical Society},
  doi = {10.1103/PRXQuantum.6.010345},
  urldate = {2026-04-10}
}

@article{leblond_logical_2025,
  title = {Logical Error Rates for the Surface Code under a Mixed Coherent and Stochastic Circuit-Level Noise Model Inspired by Trapped Ions},
  author = {LeBlond, Tyler and Groszkowski, Peter and Lietz, Justin G. and Seck, Christopher M. and Bennink, Ryan S.},
  year = 2025,
  month = nov,
  journal = {Physical Review Research},
  volume = {7},
  number = {4},
  pages = {043184},
  publisher = {American Physical Society},
  doi = {10.1103/ktb3-gcxr},
  urldate = {2026-04-17}
}

@article{lee_scalable_2025,
  title = {Scalable Projected Entangled-Pair State Representation of Random Quantum Circuit States},
  author = {Lee, Sung-Bin B. and Choi, Hee Ryang and Ohm, Daniel Donghyon and Lee, Seung-Sup B.},
  year = 2025,
  month = sep,
  journal = {Physical Review Research},
  volume = {7},
  number = {3},
  pages = {033252},
  publisher = {American Physical Society},
  doi = {10.1103/rzgm-cywf},
  urldate = {2026-05-14}
}

@article{manabe_efficient_2025,
  title = {Efficient Simulation of Leakage Errors in Quantum Error Correcting Codes Using Tensor Network Methods},
  author = {Manabe, Hidetaka and Suzuki, Yasunari and Darmawan, Andrew S},
  year = 2025,
  month = nov,
  journal = {New Journal of Physics},
  volume = {27},
  number = {11},
  pages = {114512},
  issn = {1367-2630},
  doi = {10.1088/1367-2630/ae1529},
  urldate = {2026-05-05}
}

@misc{mandelbaum_scaling_nodate,
  title = {Scaling for Quantum Advantage and beyond \textbar{} {{IBM Quantum Computing Blog}}},
  author = {Mandelbaum, Ryan},
  urldate = {2026-05-05},
  howpublished = {https://www.ibm.com/quantum/blog/qdc-2025}
}

@article{masot-llima_stabilizer_2024,
  title = {Stabilizer {{Tensor Networks}}: {{Universal Quantum Simulator}} on a {{Basis}} of {{Stabilizer States}}},
  author = {{Masot-Llima}, Sergi and {Garcia-Saez}, Artur},
  year = 2024,
  month = dec,
  journal = {Phys. Rev. Lett.},
  volume = {133},
  number = {23},
  pages = {230601},
  publisher = {American Physical Society}
}

@article{nakhl_stabilizer_2025,
  title = {Stabilizer {{Tensor Networks}} with {{Magic State Injection}}},
  author = {Nakhl, Azar C. and Harper, Ben and West, Maxwell and Dowling, Neil and Sevior, Martin and Quella, Thomas and Usman, Muhammad},
  year = 2025,
  month = may,
  journal = {Physical Review Letters},
  volume = {134},
  number = {19},
  pages = {190602},
  publisher = {American Physical Society},
  doi = {10.1103/PhysRevLett.134.190602},
  urldate = {2026-04-10}
}

@article{obrien_density-matrix_2017,
  title = {Density-Matrix Simulation of Small Surface Codes under Current and Projected Experimental Noise},
  author = {O'Brien, T. E. and Tarasinski, B. and DiCarlo, L.},
  year = 2017,
  month = sep,
  journal = {npj Quantum Information},
  volume = {3},
  number = {1},
  pages = {39},
  publisher = {Nature Publishing Group},
  issn = {2056-6387},
  doi = {10.1038/s41534-017-0039-x},
  urldate = {2026-04-17},
  copyright = {2017 The Author(s)},
  langid = {english}
}

@article{orus_practical_2014,
  title = {A {{Practical Introduction}} to {{Tensor Networks}}: {{Matrix Product States}} and {{Projected Entangled Pair States}}},
  author = {Or{\'u}s, Rom{\'a}n},
  year = 2014,
  journal = {Annals of Physics},
  volume = {349},
  pages = {117--158},
  publisher = {Elsevier},
  doi = {10.1016/j.aop.2014.06.013}
}

@misc{ott_decision-tree_2025,
  title = {Decision-Tree Decoders for General Quantum {{LDPC}} Codes},
  author = {Ott, Kai R. and Het{\'e}nyi, Bence and Beverland, Michael E.},
  year = 2025,
  publisher = {arXiv},
  doi = {10.48550/ARXIV.2502.16408},
  urldate = {2026-05-06},
  copyright = {Creative Commons Attribution 4.0 International}
}

@misc{qian_augmenting_2024,
  title = {Augmenting {{Density Matrix Renormalization Group}} with {{Clifford Circuits}}},
  author = {Qian, Xiangjian and Huang, Jiale and Qin, Mingpu},
  year = 2024,
  month = may,
  publisher = {arXiv},
  urldate = {2024-11-01}
}

@misc{quantinuum_helios_2025,
  title = {Helios: {{A}} 98-Qubit Trapped-Ion Quantum Computer},
  shorttitle = {Helios},
  author = {{Quantinuum}},
  year = 2025,
  month = nov,
  journal = {arXiv.org},
  urldate = {2026-05-05},
  howpublished = {https://arxiv.org/abs/2511.05465v1}
}

@article{seitz_simulating_2023,
  title = {Simulating Quantum Circuits Using Tree Tensor Networks},
  author = {Seitz, Philipp and Medina, Ismael and Cruz, Esther and Huang, Qunsheng and Mendl, Christian B.},
  year = 2023,
  month = mar,
  journal = {Quantum},
  volume = {7},
  pages = {964},
  publisher = {Verein zur F\"orderung des Open Access Publizierens in den Quantenwissenschaften},
  doi = {10.22331/q-2023-03-30-964},
  urldate = {2026-05-14},
  langid = {british}
}

@article{shor_scheme_1995,
  title = {Scheme for Reducing Decoherence in Quantum Computer Memory},
  author = {Shor, Peter W.},
  year = 1995,
  month = oct,
  journal = {Physical Review A},
  volume = {52},
  number = {4},
  pages = {R2493-R2496},
  issn = {1050-2947, 1094-1622},
  doi = {10.1103/PhysRevA.52.R2493},
  urldate = {2026-05-05},
  copyright = {http://link.aps.org/licenses/aps-default-license}
}

@article{sundaresan_reducing_2020,
  title = {Reducing {{Unitary}} and {{Spectator Errors}} in {{Cross Resonance}} with {{Optimized Rotary Echoes}}},
  author = {Sundaresan, Neereja and Lauer, Isaac and Pritchett, Emily and Magesan, Easwar and Jurcevic, Petar and Gambetta, Jay M.},
  year = 2020,
  month = dec,
  journal = {PRX Quantum},
  volume = {1},
  number = {2},
  pages = {020318},
  publisher = {American Physical Society},
  doi = {10.1103/PRXQuantum.1.020318},
  urldate = {2026-05-07}
}

@misc{wesdorp_mitigating_2026,
  title = {Mitigating Crosstalk Errors for Simultaneous Single-Qubit Gates on a Superconducting Quantum Processor},
  author = {Wesdorp, Jaap J. and Hyypp{\"a}, Eric and Andersson, Joona and Adam, Janos and Beriwal, Rohit and Bergholm, Ville and Dahl, Saga and Fasciati, Simone Diego and Friero, Alejandro Gomez and Gao, Zheming and Gusenkova, Daria and Guthrie, Andrew and Heinsoo, Johannes and Hiltunen, Tuukka and Holland, Keiran and Hosseinkhani, Amin and Inel, Sinan and Ikonen, Joni and Jolin, Shan W. and Juliusson, Kristinn and Kim, Seung-Goo and Komlev, Anton and Kokkoniemi, Roope and Koskinen, Otto and Kylm{\"a}l{\"a}, Joonas and Landra, Alessandro and Lamprich, Julia and Lehmuskoski, Magdalena and Lethif, Nizar and Liebermann, Per and Li, Tianyi and Lintunen, Aleksi and Marxer, Fabian and Mitra, Kunal and Mro{\.z}ek, Jakub and Ortega, Lucas and Papi{\v c}, Miha and Partanen, Matti and Plyushch, Alexander and Pogorzalek, Stefan and Renger, Michael and Ritvas, Jussi and Saarinen, Sampo and Sagar, Indrajeet and Sarsby, Matthew and Savytskyi, Mykhailo and Selinmaa, Ville and Takmakov, Ivan and Tarasinski, Brian and Tosto, Francesca and Vasey, David and Vesanen, Panu and Verjauw, Jeroen and V{\"a}limaa, Alpo and Wurz, Nicola and Ku, Hsiang-Sheng and Deppe, Frank and Hassel, Juha and {Ockeloen-Korppi}, Caspar and Liu, Wei and Tuorila, Jani and Chan, Chun Fai and Geresdi, Attila and Veps{\"a}l{\"a}inen, Antti},
  year = 2026,
  month = mar,
  number = {arXiv:2603.11018},
  eprint = {2603.11018},
  primaryclass = {quant-ph},
  publisher = {arXiv},
  doi = {10.48550/arXiv.2603.11018},
  urldate = {2026-05-13},
  archiveprefix = {arXiv}
}

@article{white_demonstration_2020,
  title = {Demonstration of Non-{{Markovian}} Process Characterisation and Control on a Quantum Processor},
  author = {White, G. A. L. and Hill, C. D. and Pollock, F. A. and Hollenberg, L. C. L. and Modi, K.},
  year = 2020,
  month = dec,
  journal = {Nature Communications},
  volume = {11},
  number = {1},
  pages = {6301},
  issn = {2041-1723},
  doi = {10.1038/s41467-020-20113-3},
  urldate = {2026-05-23},
  langid = {english}
}

@article{zhao_high-contrast_2020,
  title = {High-{{Contrast}} \${{ZZ}}\$ {{Interaction Using Superconducting Qubits}} with {{Opposite-Sign Anharmonicity}}},
  author = {Zhao, Peng and Xu, Peng and Lan, Dong and Chu, Ji and Tan, Xinsheng and Yu, Haifeng and Yu, Yang},
  year = 2020,
  month = nov,
  journal = {Physical Review Letters},
  volume = {125},
  number = {20},
  pages = {200503},
  publisher = {American Physical Society},
  doi = {10.1103/PhysRevLett.125.200503},
  urldate = {2026-05-07}
}

@misc{zhou_surface_2025,
  title = {Surface {{Code Error Correction}} with {{Crosstalk Noise}}},
  author = {Zhou, Zeyuan and Ji, Andrew and Ding, Yongshan},
  year = 2025,
  month = apr,
  number = {arXiv:2503.04642},
  eprint = {2503.04642},
  primaryclass = {quant-ph},
  publisher = {arXiv},
  doi = {10.48550/arXiv.2503.04642},
  urldate = {2025-12-17},
  archiveprefix = {arXiv}
}

\end{document}